\documentclass{article}

\usepackage{microtype}
\usepackage{graphicx}
\usepackage{subcaption}
\usepackage{booktabs} % for professional tables

\usepackage{hyperref}

\usepackage[accepted]{icml2026}

\usepackage{amsmath}
\usepackage{amssymb}
\usepackage{mathtools}
\usepackage{amsthm}
\usepackage{fontawesome}

\usepackage[capitalize,noabbrev]{cleveref}

\theoremstyle{plain}

\theoremstyle{definition}

\theoremstyle{remark}

\usepackage[textsize=tiny]{todonotes}

\icmltitlerunning{Interpretable Neural Marked Statistics for Cosmological Inference}

\usepackage{colortbl}
\definecolor{linkcolor}{rgb}{0.0,0.3,0.5}

\definecolor{darkred}{RGB}{175,0,0}
\definecolor{darkblue}{RGB}{14,0,185}
\definecolor{salmon}{RGB}{255,160,105}
\definecolor{redd}{RGB}{0,150,0}

\definecolor{darkteal}{RGB}{0,111,111}

\newcommand{\bx}{\mathbf{x}}
\newcommand{\bk}{\mathbf{k}}

\newcommand{\bz}{\mathbf{z}}

\newcommand{\bS}{\mathbf{S}}
\newcommand{\btheta}{\boldsymbol{\theta}}

\newcommand{\kmax}{k_{\mathrm{max}}}
\newcommand{\kn}{k_{\mathrm{Nyq}}}
\newcommand{\Pdd}{P_{\delta\delta}}
\newcommand{\Pdm}{P_{\delta\Delta}}
\newcommand{\Pmm}{P_{\Delta\Delta}}

\begin{document}

\twocolumn[
  \icmltitle{Interpretable Equivariant Marks for Contrastive Cosmological Inference}



  \icmlsetsymbol{equal}{*}

  \begin{icmlauthorlist}
    \icmlauthor{Federico Semenzato}{unipd,infn}
    \icmlauthor{Benjamin D. Wandelt}{jhu1,jhu2}
    \icmlauthor{Michele Liguori}{unipd,infn,inaf}
    \icmlauthor{Alvise Raccanelli}{unipd,infn,inaf}
  \end{icmlauthorlist}

\icmlaffiliation{unipd}{Dipartimento di Fisica Galileo Galilei, Universit\`a di Padova, I-35131 Padova, Italy}
\icmlaffiliation{infn}{INFN Sezione di Padova, I-35131 Padova, Italy}
\icmlaffiliation{inaf}{INAF-Osservatorio Astronomico di Padova, Italy}
\icmlaffiliation{jhu1}{Department of Physics and Astronomy, Johns Hopkins University, 3400 North Charles Street, Baltimore, MD, 21218, USA}
\icmlaffiliation{jhu2}{Department of Applied Mathematics and Statistics, Johns Hopkins University, 3400 North Charles Street, Baltimore, MD, 21218, USA}

  \icmlcorrespondingauthor{Federico Semenzato}{federico.semenzato.1@phd.unipd.it}

  \icmlkeywords{Machine Learning, ICML}

  \vskip 0.3in
]



\printAffiliationsAndNotice{}  % no special notice (required even if empty)

\begin{abstract}
Recovering cosmological information beyond the power spectrum is a central goal for upcoming cosmological surveys, since late-time non-Gaussian signal in the matter density cannot be accessed through two-point statistics alone.
Marked statistics fold part of this information back into the two-point level by reweighting the field with non-linear functions. 
We propose a neural marking scheme to generalize this process through a set of interpretable, physically motivated transformations that directly allow to interpret the gain in cosmological information at the morphological level.
We employ a contrastive learning objective to align learnable marked summaries with the underlying cosmological parameters.
At~$k_{\max}=0.2\,h\mathrm{Mpc}^{-1}$, our neural mark tightens the marginalized constraint on~$\sigma_8$ by~$2.9\times$ and on~$\Omega_m$ by~$1.8\times$ compared to classical marks, breaking the~$\Omega_m-\sigma_8$ degeneracy at the Fisher information level. It further reduces the parameter MSE across our cosmological parameter prior by~$1.45\times$ over the best classical mark.
The learned latent geometry aligns with the~$\Omega_m$ and~$\sigma_8$ directions in parameter space, indicating that the contrastive objective recovers the dominant axes of cosmological information.
Our approach opens the door to more powerful, interpretable summary statistics for cosmological inference.
\end{abstract}

\section{Introduction}
\label{sec:intro}

Large-scale structure (LSS) surveys provide three-dimensional maps of millions of galaxies, with the current and next generation of surveys expanding the scale and precision of these maps (e.g.,~SPHEREx~\cite{Dore:2014cca}, DESI~\cite{desicollaboration2016desi}, Euclid~\cite{euclidcollaboration2024euclid}).
Their statistical analysis  can in principle constrain some of the deepest questions in fundamental physics: the amplitude of primordial non-Gaussianity and the inflationary models it discriminates~\cite{sabino_png,maldacena:2003,Dalal_2008,matarrese_effect_2008}, the dark energy equation of state~\cite{DESI:2024hhd}, and the absolute neutrino mass scale~\cite{Lesgourgues:2006nd,Elbers:2025vlz}.
Much of this signal lives in correlations of the late time matter density that the Gaussian two-point structure alone cannot capture~\cite{Bernardeau:2001qr,Scoccimarro:1999ed,Sefusatti:2007fn}; extracting it is crucial for the scientific return of these surveys.

\paragraph{Two-point summaries break in the non-linear regime.} The power spectrum is the canonical summary of  LSS clustering: optimal for Gaussian fields,  modelable with perturbation theory on quasi-linear scales, and well understood at the level of covariance and systematics~\cite{Tegmark_1997,Percival_2004,DESI:2024mwx}.
The information beyond the linear regime lives in higher-order correlations~\cite{Scoccimarro:1999ed,verde/etal:2000, Bernardeau:2001qr, Bertacca:2014wga} but climbing the ladder of~$n$-point correlators directly is expensive in configuration space, perturbation theory breaks down in the non-linear regime, and the Gaussian-likelihood assumptions underlying the choice of these summaries become increasingly fragile.

\paragraph{Field-level and hybrid neural summaries.} Simulation-based inference and field-level neural summaries sidestep explicit summaries by leveraging simulated data and mapping them directly to parameter posteriors~\cite {imnn,Makinen:2021nly,SimBIG:2023ywd}.
These approaches require many high-fidelity simulations, and trade constraining power for interpretability of the features that drive  cosmological parameter constraints.
Hybrid strategies blend neural summaries with traditional statistics,  capturing information beyond analytically tractable summaries while leaving the neural component opaque~\cite{Makinen:2024xph, Bairagi:2025ytq}.

\paragraph{Marked statistics.} A complementary route reweights the field by a spatial mark~$M(\bx)$; the two-point spectra of the resulting marked field then mix higher-order correlators of the original field~\cite{Philcox_2020_pert,Marinucci:2024bdq,Ebina:2026qzf}.
The summary stays a power spectrum, while the choice of mark sets which higher-order structure is folded down to the two-point level~\cite{White:2016yhs,Massara:2020pli}.
Existing constructions fix this transformation from a smoothed-density response~\cite{White:2016yhs,Massara:2020pli,Ebina:2024zkv}.
Recent work extends classical mark functions by optimizing their functional shape for a fixed smoothing scale~$R$ through Fisher information maximization~\citep{Cowell:2024wyl} at the fiducial cosmology. Both restrict the mark to narrowly parametrized forms and keep it tied to a single point in cosmological parameter space.

\paragraph{Our contribution.} 
We replace the hand-designed mark with a learned, cosmology-agnostic one that preserves  interpretability through physically motivated architectural constraints. The mark is built from three, locally SO(3)-equivariant spherical harmonic filters ($\ell\!\in\!\{0,1,2\}$) reduced to four rotation-invariant scalar channels and processed by independent MLPs that additively combine into the mark function. 
Rather than assuming which environmental features matter the most, we can learn which aspects of the local cosmic web geometry carry the most cosmological information.
The result is a trained marking function that remains inspectable in configuration space and encodes which local morphological features carry the richest cosmological signal.

To search this enlarged mark space, we borrow a tool from multimodal representation learning.  We align marked summaries with cosmological parameters ~$\btheta$ in a shared latent space using a contrastive InfoNCE objective~\cite{Oord_2018,Radford:2021,Chen_2020} with a learned Mahalanobis metric.
We residualize the marked-summary embedding against the unmarked embedding so the mark is rewarded only for complementary information. On Quijote N-body simulations, the resulting summary tightens marginalized Fisher constraints by~$2.9\times$ on~$\sigma_8$ and~$1.8\times$ on~$\Omega_m$ at~$k_{\max}=0.2\,h\mathrm{Mpc}^{-1}$, breaks the canonical~$\Omega_m$--$\sigma_8$ degeneracy, and reduces held-out parameter MSE across the parameter space by up to~$\sim\!1.45\times$ over the best classical marks.

\section{Methodology}
Our framework is built of two main components.
A learnable mark architecture extracts rotation-invariant local descriptors of the density field (\cref{sec:learned_marks}), and a contrastive objective aligns marked summaries to cosmological parameters in a shared latent space (\cref{sec:contrastive}).
We introduce an explicit residualization of the marked-statistic embeddings against the unmarked-statistic embeddings so the mark is rewarded only for information it adds beyond the standard power spectrum.

\subsection{Marked Two-Point Summaries}
\label{sec:marked_summaries}
We define a mark field~$M(\bx)$ as a scalar function over the simulation volume, which weights different regions of the matter density contrast ~$\delta(\bx)$ according to their local properties. 
The (mean-centered) marked density field then reads
\begin{equation}
  \Delta(\bx)
  =
  M(\bx)\,[1+\delta(\bx)]
  -
  \left\langle M(1+\delta)\right\rangle   \,.
  \label{eq:marked_field}
\end{equation}
To map the marked field to cosmological parameters, we use the set of two-point power spectra of the marked and unmarked fields, that is
\begin{equation}
  \bS(k)=
  \left\{\Pdd(k),\ \Pdm(k),\ \Pmm(k)\right\}\,,
  \label{eq:spectra_set}
\end{equation}
\begin{equation}
  \langle a(\bk)b^\ast(\bk')\rangle
  =(2\pi)^3\delta_D(\bk-\bk')P_{ab}(k)\,,
  \label{eq:marked_spectra}
\end{equation}
with~$a,b\in\{\delta,\Delta\}$. Because~$\Delta$ is nonlinear in~$\delta$,~$\Pmm$ and~$\Pdm$ include higher-order correlations of the original field~\citep{Philcox_2020_pert,Marinucci:2024bdq}.
The non-linear transformation~$M(\delta_R)$ effectively folds higher-order clustering into a modified two-point function.
Crucially,~\citep{Cowell:2024wyl} showed that the constraints on cosmological parameters derived from the combination of auto- and cross-spectra~$\{P_{\delta\delta}, P_{\delta\Delta}, P_{\Delta\Delta}\}$ are invariant for affine transformations of the mark~$M' \to aM+b$.

Classical marks usually depend on a smoothed density~$\delta_R$,
\begin{equation}
  M(\bx)=
  \left(
    \frac{1+\delta_s}{1+\delta_s+\delta_R(\bx)}
  \right)^p ,
  \label{eq:classical_mark}
\end{equation}
with smoothing scale~$R$ and saturation parameter~$\delta_s$~\citep{White:2016yhs,Massara:2020pli}.
Positive~$p$ up-weights underdense regions; negative~$p$ emphasizes overdense or nonlinear regions.
Learned marks expand the transformation space while preserving the two-point output in \cref{eq:marked_spectra}.

\begin{figure*}[t]
  \centering
  \begin{subfigure}[t]{0.48\textwidth}
    \centering
    \includegraphics[width=\linewidth]{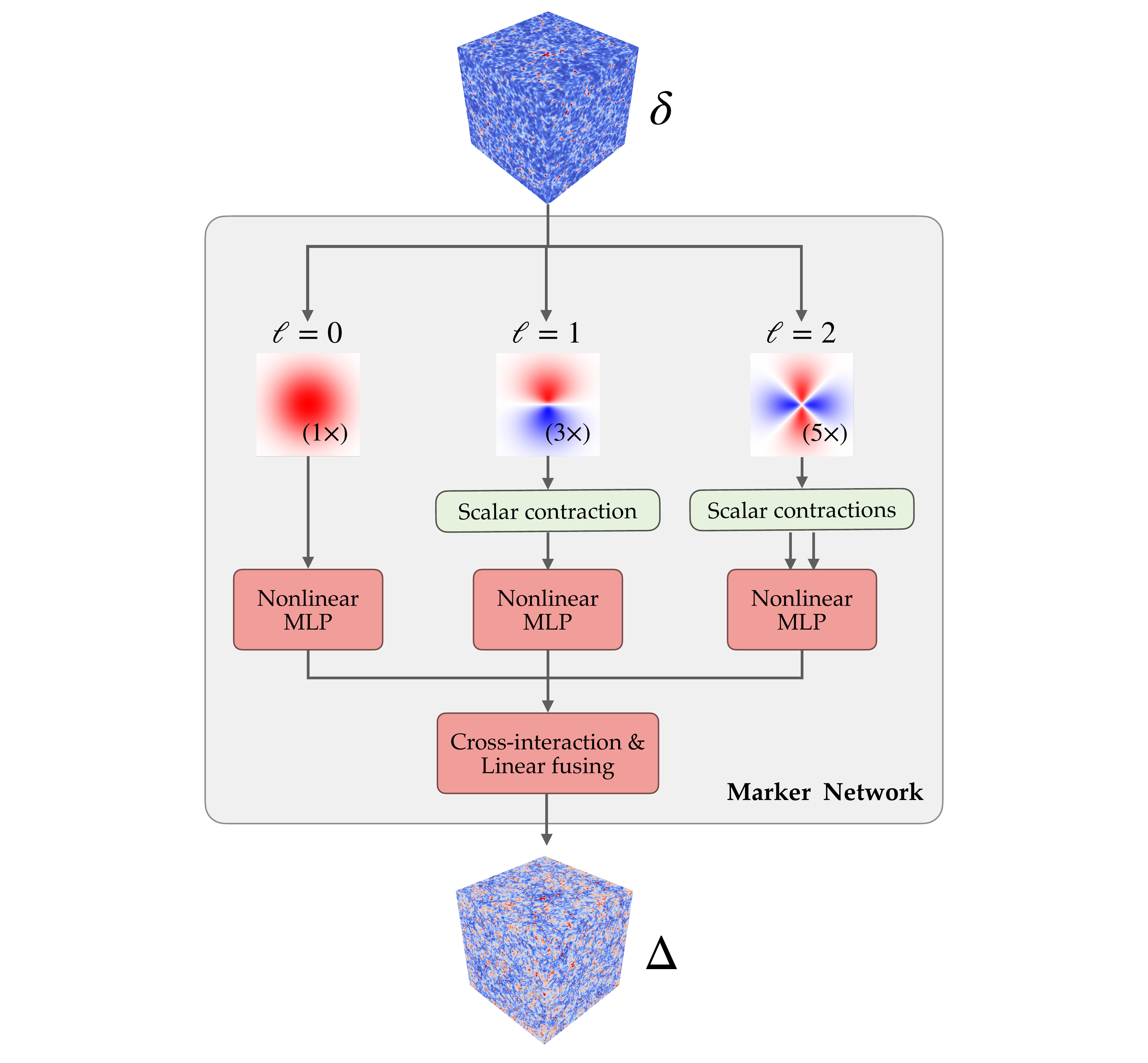}
    \caption{Interpretable mark module.}
    \label{fig:method_overview_a}
  \end{subfigure}
  \hfill
  \begin{subfigure}[t]{0.48\textwidth}
    \centering
    \includegraphics[width=\linewidth]{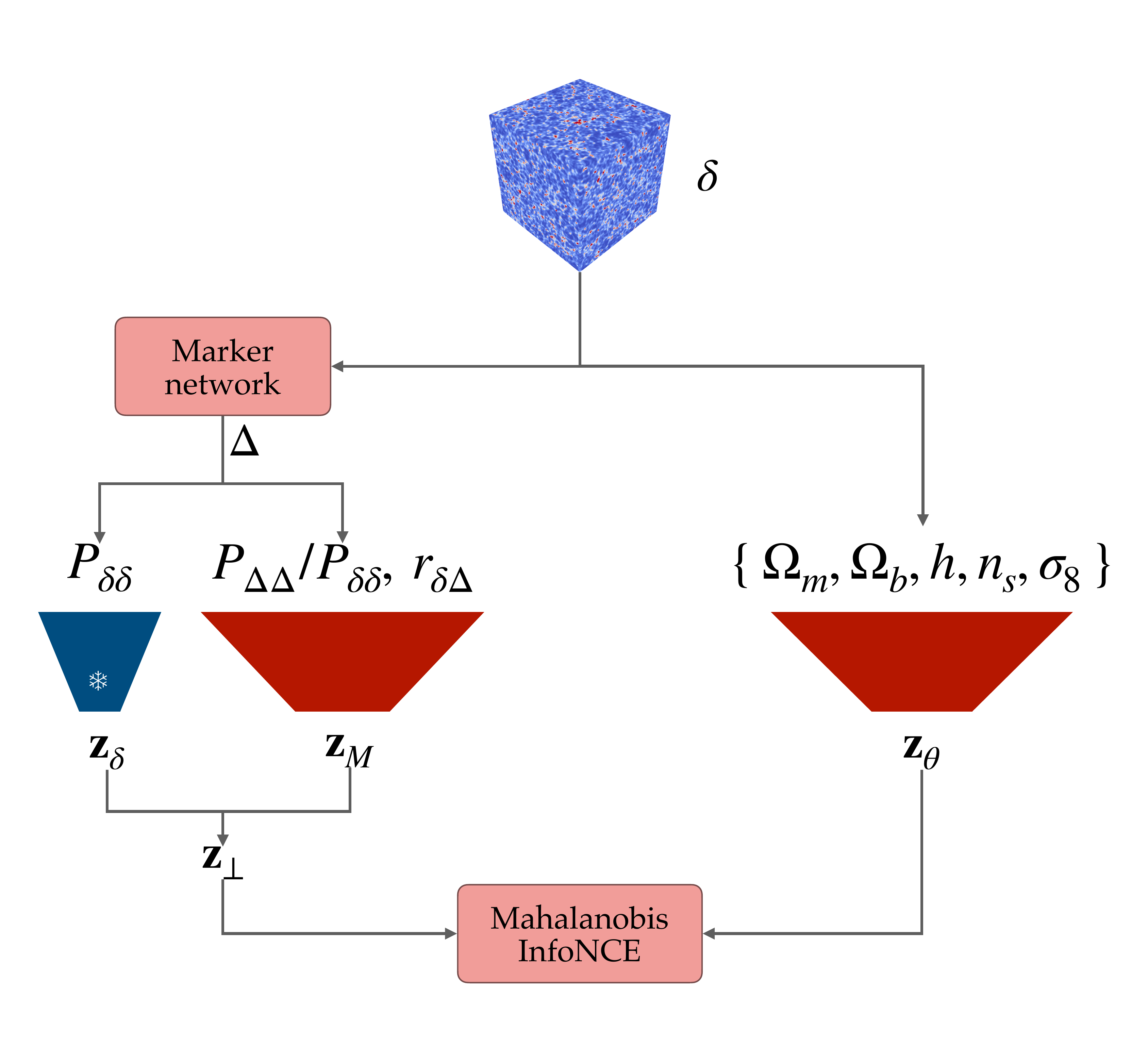}
    \caption{Summary alignment and contrastive setup.}
    \label{fig:method_overview_b}
  \end{subfigure}
  \caption{The learned mark aims to extract interpretable features from the density field.
(a) Spherical-harmonic filters~$(\ell=0,1,2)$ produce density-like, gradient-like, and quadrupole-like responses. Scalar contractions form rotation invariant fields which are independently processed by MLPs and linearly combined with a small cross interaction. The resulting mark~$M$ produces a marked field~$\Delta$.
(b) The~$P_{\delta\delta}$ embedder is pretrained against parameters and frozen; the marked summary embedding~$\bz_M$ is residualized against~$\bz_\delta$ to give~$\bz_\bot$, which is aligned with~$\bz_\theta$ through a learned Mahalanobis-form InfoNCE loss.}
  \label{fig:method_overview}
\end{figure*}

\subsection{Interpretable Learned Marks}
\label{sec:learned_marks}
We make use of matter density fields at ~$z=0$  produced by the Quijote simulation suite of N-body simulations~\citep{Villaescusa_Navarro_2018}.
Particles are placed in periodic boxes of side length~$1~h^{-1}\mathrm{Gpc}$, and assigned to~$128^3$ grids.
The mark is therefore a translation-equivariant local rule applied to every voxel.

We decompose the density field into spherical-harmonic-aware filtered channels and then combine them into scalar invariants.
The filtered maps are
\begin{equation}
  \begin{aligned}
  F_{\ell m}(\bx)
  =
  \mathcal{F}^{-1}\!\big[
    &\widetilde{\delta}(\bk)
    W_{\mathrm{MAS}}^{-1}(\bk)
    T_{\mathrm{tap}}(k)\\
    &{}\times G_\ell(k)
    i^\ell Y_{\ell m}(\hat{\bk})
  \big](\bx),
  \end{aligned}
  \label{eq:filtered_fields}
\end{equation}
for~$\ell\le 2$.
Here~$G_\ell(k)$ is a learned radial profile in Fourier space and~$Y_{\ell m}$ is a real spherical harmonic.
The factor~$W_{\mathrm{MAS}}^{-1}$ removes the mass-assignment window used to place particles on the grid~\cite{Hand_2018}.
The spherical Nyquist cutoff~$T_{\mathrm{tap}}(k)$ suppresses high-$k$ modes amplified by the deconvolution that cannot be reliably modeled because of aliasing, with a mild Tukey cosine roll-off to control cubic-grid artifacts.
Each~$G_\ell(k)$ is parameterised as a learnable Gaussian band-pass with shared center~$r_0$ and width~$\sigma$ plus a per-$\ell$ residual MLP on log-spaced Fourier features, zero-initialized \citep{2020arXiv200308934M}.
%Each~$G_\ell(k)$ is parametrized as a learnable linear combination of a fixed set of Gaussians equally spaced in~$k$ with shared width, so the radial profile reduces to one coefficient vector per~$\ell$, zero-initialized.

The~$\ell=0$ block is a signed monopole response,~$E_0=F_{00}$, while the~$\ell=1$ block gives a vector-like response whose invariant amplitude is
\begin{equation}
  E_1=\left(\sum_{m=-1}^{1}F_{1m}^2+\epsilon\right)^{1/2}.
  \label{eq:e1}
\end{equation}
For~$\ell=2$, the five real spherical-harmonic components~$F_{2  m}$ map to the five independent entries of a traceless symmetric tensor~$Q$, from which we extract two natural scalar invariants:
\begin{equation}
  E_2=\left[\mathrm{Tr}(Q^2)+\epsilon\right]^{1/2}\,,
  \qquad
  I_3=\frac{\mathrm{Tr}(Q^3)}{E_2^3+\epsilon}\,.
  \label{eq:quadrupole_invariants}
\end{equation}
$E_2$ measures anisotropy strength, while~$I_3$ distinguishes prolate and oblate quadrupole shape.

The invariant channels are processed by independent scalar MLPs,
\begin{align}
  M(\bx) &= \mathrm{softplus}\,[\eta(\bx)],\\
  \eta(\bx)
  &=
  \sum_a h_a(f_a(\bx))\nonumber\\
  &\quad
  + h_{\times}\!\left(E_0(\bx),E_2(\bx),I_3(\bx)\right),
  \label{eq:gam_mark}
\end{align}
where~$f_a\in\{E_0,E_1,E_2,I_3\}$.
The cross term~$h_{\times}(E_0,E_2,I_3)$ couples density to tidal-like features~$E_2$ and~$I_3$ to provide a probe of how density modulates anisotropic structure. Richer cross-interactions and full equivariant mixing of the~$\ell$-channels are natural extensions we leave to future work.

The output is initialized at~$M(\bx)\simeq1$, so training begins at the unmarked field.
Because each~$h_a$ is separate, the learned mark can be decomposed into distinct contributions before spectra are evaluated.

\begin{figure*}[t]
  \centering
  \includegraphics[width=\textwidth]{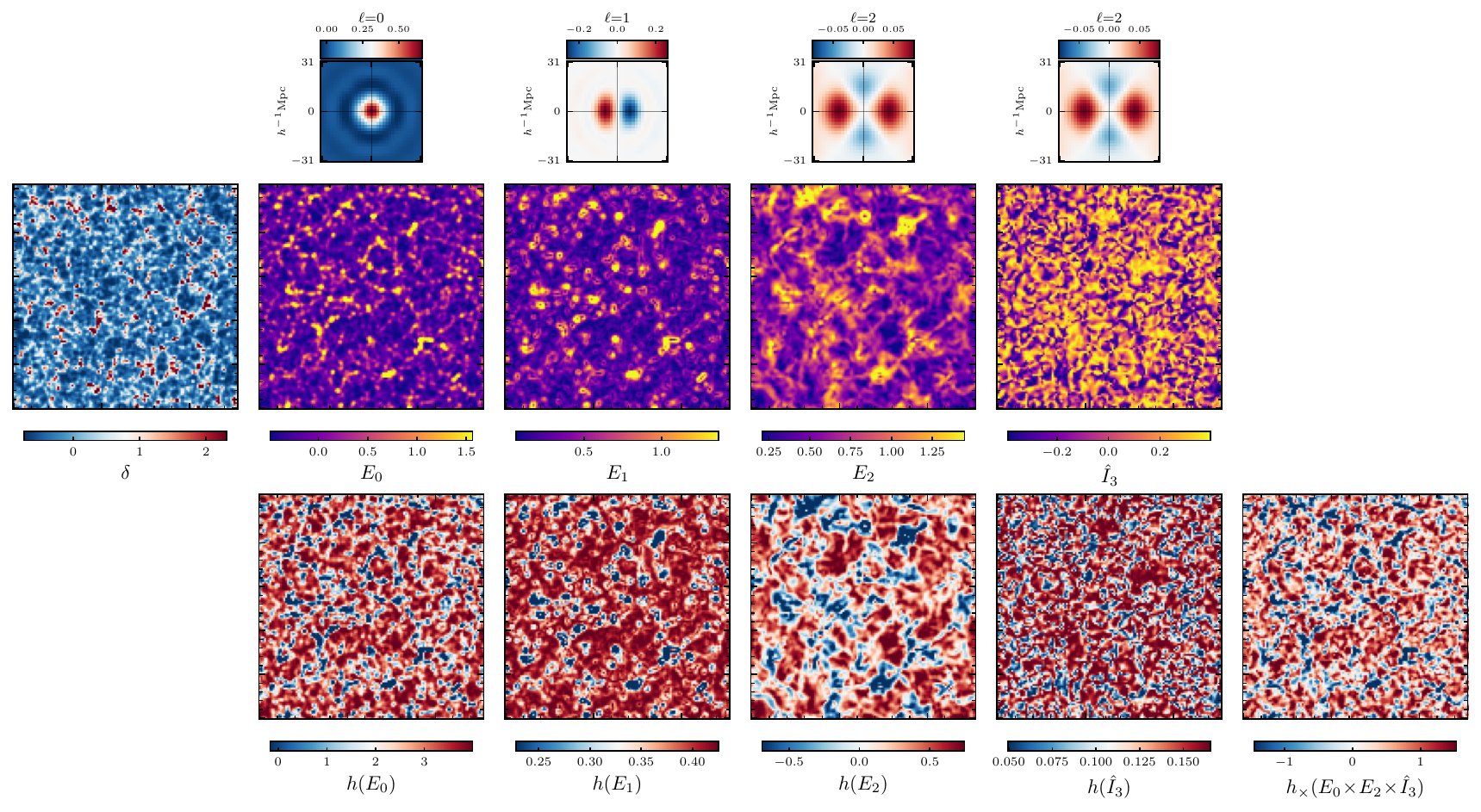}
  \caption{End-to-end introspection of the learned mark on a representative held-out simulation.
\textbf{Top row}: real-space projection of the per-$\ell$ kernels reconstructed from the trained~$G_\ell(k)$, for~$\ell=0,1,2$.
\textbf{Middle row}: the input density field~$\delta$ and the four rotation-invariant maps~$E_0,E_1,E_2,I_3$ extracted from the spherical-harmonic-filtered field.
\textbf{Bottom row}: the learned scalar response contributions~$h_a(E_0)$,~$h_a(E_1)$,~$h_a(E_2)$,~$h_a(I_3)$, and the cross-interaction term~$h_\times(E_0,E_2,I_3)$ that combine into the mark~$M$.
The isotropic ($\ell=0$) channel dominates the response on these scales; the anisotropic channels are present but less dominant.
}
  \label{fig:introspection_main}
\end{figure*}

\subsection{Contrastive Training}
\label{sec:contrastive}
The contrastive objective operates on the summary vectors for each simulation.
The unmarked summary is
\begin{equation}
  \bS_\delta=\log_{10}\Pdd(k)\,,
\end{equation}
while the marked summary is built as
\begin{align}
  \bS_M(k)
  &=
  \left[
    \log_{10}\frac{\Pmm(k)}{\Pdd(k)},\
    r(k)
  \right],\\
  r(k)
  &=
  \frac{\Pdm(k)}{\sqrt{\Pdd(k)\Pmm(k)}} .
  \label{eq:summary_features}
\end{align}
\nobreak

For simulation~$i$, the positive pair is therefore~$(\bS_{M,i},\btheta_i)$.
We project the two summary vectors and the cosmological parameters into a shared~$D$-dimensional embedding space, where the contrastive loss is applied.
The parameter projector is chosen to be linear.

In latent space, we remove the component of the marked embedding parallel to the unmarked embedding,
\begin{equation}
  \bz_{\perp}
  =
  \bz_M
  -
  \frac{\bz_M\cdot\bz_\delta}{\bz_\delta\cdot\bz_\delta+\epsilon}\,
  \bz_\delta .
  \label{eq:embedding_ortho}
\end{equation}
This projection discourages the learned mark from simply reproducing information already available in~$\Pdd$.
The contrastive loss then aligns each marked summary embedding~$\bz_{\perp,i}$ with the corresponding parameter embedding vector~$\bz_{\theta,i}$, while separating it from a set~$\mathcal{N}_i$ of specific negative candidates~$\bz_{\theta,j}$.
We use an InfoNCE loss~\cite{Oord_2018}:
\begin{equation}
  \mathcal{L}_i
  =
  -\log
  \frac{\exp s(\bz_{\perp,i},\bz_{\theta,i})}
  {\sum_{j\in\mathcal{N}_i^+}\exp s(\bz_{\perp,i},\bz_{\theta,j})},
  \label{eq:contrastive_loss}
\end{equation}
where~$\mathcal{N}_i^+$ includes both the negatives~$\mathcal{N}_i$ and the true parameter embedding.
Since generating~$\bz_{\theta,j}$ only requires a cosmological parameter vector without the respective~$\bz_{\perp,j}$ (and thus a paired density field), we can efficiently design  the negative sampling scheme.
We extract these contrastive negatives from three regions of the Quijote prior: real negatives are drawn from the training set and shared across elements of a batch; global synthetic negatives sample the full prior volume; and local synthetic negatives sample a shell around the anchor cosmology.
The local candidates force the summary to distinguish nearby cosmologies without requiring additional simulations.
The score is a  Mahalanobis distance,
\begin{equation}
  s(\bz_{\perp},\bz_\theta)
  =
  -\tau^{-1}
  (\bz_{\perp}-\bz_\theta)^TLL^T(\bz_{\perp}-\bz_\theta),
  \label{eq:mahalanobis}
\end{equation}
with a learned lower-triangular factor~$L$ and fixed temperature~$\tau$.
This metric allows rotations and stretchings of latent space to better align the summary geometry with the parameter geometry, while still penalizing large distances in any direction.

We first pretrain the unmarked branch to align~$\Pdd$ with parameter embedding. 
We then freeze the ~$\Pdd$-embedder weights when training the marker module. 
The parameter embedder is initialized to the configuration learned for~$\Pdd$ and allowed to fine-tune during mark training, so the final geometry is not fixed by the unmarked summary.
Without this bootstrap, the unmarked embedder could collapse and make the complementarity constraint meaningless.
Loss-function hyperparameters and optimizer settings are reported in \cref{app:hp_optim}.

% It therefore has enough flexibility to represent the anisotropic geometry of cosmological degeneracies, including the~$\Omega_m$--$\sigma_8$ direction, without forcing all parameter sensitivity to align with coordinate axes.

\section{Experiments}
\label{sec:experiments}
We train our contrastive framework on a set of~$5,000$ simulations from the Quijote BSQ suite~\cite {Bairagi:2025sux}. 
The dataset comprises simulations with~$5$ varying cosmological parameters~$\btheta=(\Omega_m,\Omega_b,h,n_s,\sigma_8)$, with~$\Omega_m\in[0.1,0.5]$,~$\Omega_b\in[0.02,0.08]$,~$h\in[0.5,0.9]$,~$n_s\in[0.8,1.2]$, and~$\sigma_8\in[0.6,1.0]$.
Density fields are assigned to~$128^3$ grids, giving a cell size of~$7.8~h^{-1}\mathrm{Mpc}$ and~$\kn=\simeq0.4~h\,\mathrm{Mpc}^{-1}$.
The contrastive training uses spectra through~$k\simeq0.3~h\,\mathrm{Mpc}^{-1}$, to avoid artifacts close to~$\kn$.
Once the contrastive training has converged, the learned mark is a fixed, interpretable transform of the density field.
We can then evaluate the per-channel response inside the mark itself, and extract marked spectra for regular downstream cosmological constraints.
The code developed for this analysis can be found \href{https://github.com/fsemenzato/Interpretable-Equivariant-Marks}{here \faGithub}.

\subsection{Mark Introspection}
\label{sec:introspection}

\cref{fig:introspection_main} traces the trained mark from the input density down to the per-channel scalar contributions that combine to give~$M(\bx)$.
The top row shows the per-$\ell$ kernels reconstructed from the trained~$G_\ell(k)$: the monopole is broad and approximately isotropic, the dipole resembles a derivative-of-Gaussian profile, and the quadrupole exhibits the expected oriented six-lobe pattern.
The middle row shows the bare density~$\delta$ alongside the rotation invariants~$E_0$,~$E_1$,~$E_2$ and the cubic~$I_3$:~$E_1$ peaks at the boundaries between voids and filaments,~$E_2$ in elongated filamentary regions, and~$I_3$ separates oblate from prolate quadrupole configurations at fixed~$E_2$.
The bottom row shows the per-channel scalar responses~$h_a(E_a)$ together with the cubic cross-interaction~$h_\times(E_0,E_2,I_3)$, which sum (modulo the output activation) to the scalar mark~$M$.
At this resolution and~$\kmax$, the response is dominated by the isotropic~$E_0$ channel: the trained~$G_0(k)$ acts as a high-pass filter, so the network effectively recovers a non-linearly reweighted version of the density. The dipole and quadrupole responses contribute subdominant corrections at structure boundaries and in elongated regions, and~$I_3$ adds only a weak modulation.
In \cref{app:kmax_ablation} we quantify the relative contribution of the anisotropic channels in terms of Fisher constraints, comparing the full mark to an isotropic~$\ell_{\max}=0$ ablation.
$E_0$ carries the largest fraction of the gain in mrginalized confidence level, with the anisotropic channels refining the constraints in this setup.
The additive form of \cref{eq:gam_mark} makes this channel decomposition exact rather than a saliency proxy: the panels in the bottom row are literally the terms that combine into~$M(\bx)$, so the dominance of~$E_0$ is a property of the trained mark itself, not an artifact of visualization. The same observation motivates the morphology-aware extensions discussed in \cref{sec:conclusion}.

\subsection{Summary-Level Constraints}

\begin{figure}[t]
  \centering
  \includegraphics[width=0.88\columnwidth]{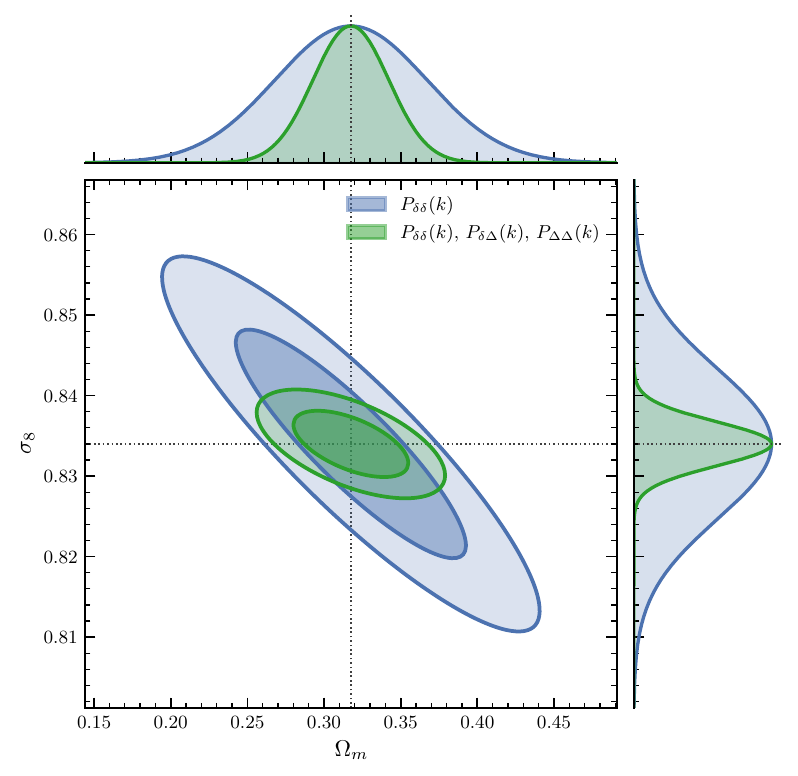}
  \caption{Marginalized posterior contours from the Fisher information matrix at the fiducial cosmology, for~$\Pdd$ alone (blue) and the learned mark~$\{\Pdd,\Pdm,\Pmm\}$ (green). The learned mark tightens the one-dimensional posteriors on every parameter and rotates the~$\Omega_m$--$\sigma_8$ contour, breaking the canonical degeneracy and enabling simultaneous improvement on both parameters.}
  \label{fig:fisher_triangle}
\end{figure}

\label{sec:constraints}
\begin{figure}[t]
  \centering
  \includegraphics[width=0.88\columnwidth]{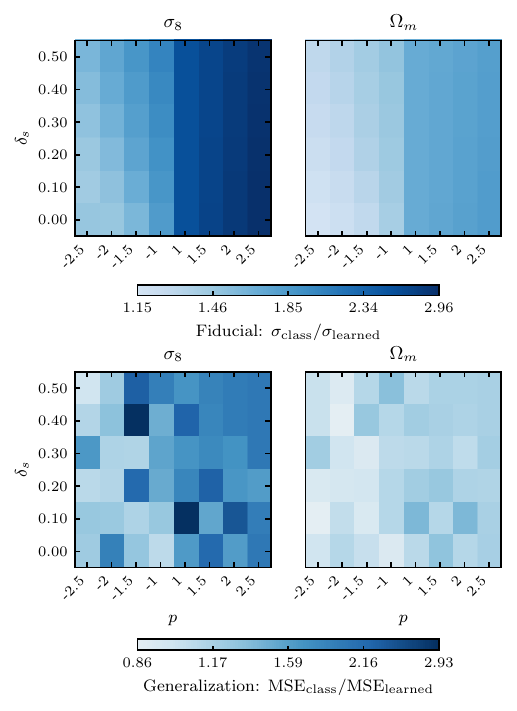}
  \caption{Comparison of learned and classical marks on both Fisher and held-out generalization at~$\kmax\!=\!0.20\,h\,\mathrm{Mpc}^{-1}$.
\textbf{Top}: Improvement in the marginalized confidence intervals relative to a set of classical marks from~\cite {Massara:2020pli} with~$R=10~h^{-1}\mathrm{Mpc}$.
\textbf{Bottom}: Mean-squared error from a MLP regressor on Latin-hypercube simulations unseen during training; the learned mark beats the unmarked baseline and the best classical mark, with the largest gains on~$\sigma_8$.}
  \label{fig:results_compact}
\end{figure}

We evaluate the learned summary on two complementary tasks at~$\kmax\!=\!0.20\,h\,\mathrm{Mpc}^{-1}$, well inside the regime where mass-assignment artifacts are negligible and the learned mark is active.
At higher~$k_{\max} \geq 0.3\,h\mathrm{Mpc}^{-1}$, the anti-aliasing taper required by the spherical-harmonic filtering on the cubic grid removes power that classical pixel-space marks retain, and some classical marks become competitive (see \cref{app:kmax_ablation}). Extensions of this framework to morphology-based filters, which we will report in follow-up work, restore uniform superiority.

The Fisher information matrix~\cite{Tegmark_1997,Heavens_1995} provides the standard forecasting tool at the fiducial cosmology.
Under a Gaussian likelihood with covariance~$C$ and parameter-dependent mean~$\mu(\btheta)$,~$F_{\alpha\beta}=\nabla_\alpha\mu^T C^{-1}\nabla_\beta\mu$, and the Cram\'er--Rao inequality bounds the achievable variance on any unbiased estimator by~$F^{-1}$.
We compute the covariance by evaluating the spectra on~$10,000$ fiducial simulations and the derivatives by finite differences on~$500$ simulations per parameter.
These fiducial and derivative simulations are disjoint from the BSQ set used for contrastive training and from the Latin-hypercube set used for the held-out tests.

In \cref{fig:fisher_triangle} we show the marginalized posterior contours for~$\Pdd$ alone and the learned mark summaries~$\{\Pdd,\Pdm,\Pmm\}$.
The learned mark tightens the one-dimensional posteriors on every parameter.
Importantly, the additional marked spectra also allow the contour to rotate, breaking the canonical~$\Omega_m\hbox{--}\sigma_8$ degeneracy and enabling simultaneous improvement on both parameters.
In the top row of \cref{fig:results_compact} we compare the improvement in the marginalized confidence intervals relative to a set of classical marks from~\cite {Massara:2020pli} with~$R=10~h^{-1}\mathrm{Mpc}$, which includes the best mark parameters found in that work. The learned mark improves on the best classical mark for every parameter, with the largest gains on~$\sigma_8$ and~$\Omega_m$.

On the other hand, with a further held-out test set spanning the entire Hypercube, we train an MLP from each summary to the five-parameter vector to evaluate generalization performance across the full parameter volume. 
In \cref{app:appendix_mlp} we provide details of the MLP architecture and training procedure.
In the bottom row of \cref{fig:results_compact} we show the mean-squared error (MSE) from this regressor on Latin-hypercube simulations unseen during training.
The learned mark presents good generalization performance with respect to the configurations tested, with particular gains on~$\sigma_8$.
For a small number of~$(p,\delta_s)$ configurations, the per-parameter MSE ratio dips marginally below unity on~$\Omega_m$, so the best classical marks remain slightly more stable on~$\Omega_m$. 
Conversely, the learned density modulation efficiently generalizes on~$\sigma_8$ across all configurations tested.

%\cref{fig:results_compact} reports both side by side.
Adding the marked spectra to~$P_{\delta\delta}$ tightens the marginalized Fisher posteriors on every cosmological parameter and 
the learned summary reaches a lower error than any classical~$(p,\delta_s,R)$ choice on the same grid, with the largest gains on~$\sigma_8$ and~$\Omega_m$.
The held-out MLP delivers the same conclusion in the generalization regime.
Across both tasks, the learned mark outperforms classical mark configurations in this setting.

\subsection{Latent-Space Geometry}
\label{sec:latent}
\begin{figure}[t]
  \centering
  \includegraphics[width=\linewidth]{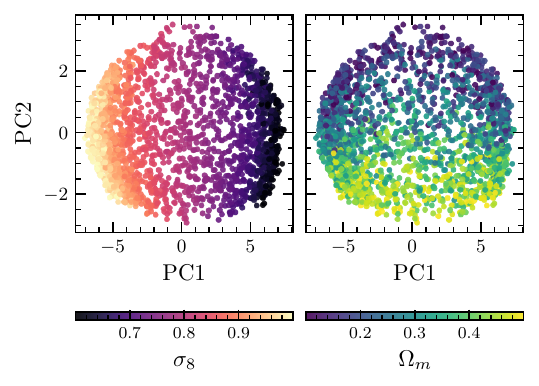}
  \caption{Two-dimensional PCA projection of the complementary query embedding~$\bz_{\perp}$ for held-out simulations, colored by~$\sigma_8$ (left) and~$\Omega_m$ (right).~$\sigma_8$ varies smoothly along the first principal direction and~$\Omega_m$ along an approximately orthogonal one. The first two PCs therefore align with the most well-constrained parameters of the problem.}
    \label{fig:latent_main}
\end{figure}

We further explore the learned geometry of the marked summary embedding~$\bz_{\perp}$.
The contrastive objective shapes the marked-summary embedding~$\bz_{\perp}$ into a~$D=16$ space, but the trained representation has effective rank~$\approx 3.4$, with the first two principal components carrying~$\sim85\%$ of the variance.
Most of the signal therefore lives on a small subspace, which is consistent with the intrinsic dimensionality of the cosmological-parameter target.

\cref{fig:latent_main} shows that the first two principal components of~$\bz_{\perp}$ on held-out simulations align almost perfectly with~$\sigma_8$ and~$\Omega_m$, exactly the two parameters that dominate the constraint gain in \cref{fig:results_compact}.
This alignment is a representation-level signature consistent with the breaking of the~$\Omega_m\hbox{--}\sigma_8$  degeneracy at the parameter constraint level.

Using the same held-out simulations as in~\cref{sec:constraints}, we evaluate the fraction~$R@k$ of simulations whose true parameters fall among the~$k$ nearest candidates.
%Under the learned metric on~$2000$ Latin-hypercube simulations,~$R@1=66.5\%$ and~$R@5=96.1\%$ for the orthogonalized query embedding~$\bz_{\perp}$.
For each of the ~$2000$ Latin-hypercube simulations~$i$, we compute the metric distance (see \cref{eq:mahalanobis}) between the orthogonalized mark embedding~$\bz_{q,i}$ and all the other parameter embeddings~$\bz_{\theta,j}$ in the set, with the matched index~$j\!=\!i$ as the only correct candidate.
In terms of recall-at-$k$, we obtain~$R@1\!=\!66.5\%$,~$R@5\!=\!96.1\%$, against a random baseline~$R@1\!=\!1/N\!=\!0.05\%$.
This test does not include the intrinsic cosmic-variance scatter of the summaries, which sets a floor on the achievable retrieval performance from a single realization.
Unlike the Fisher forecast in \cref{sec:constraints}, which measures \emph{local} sensitivity at the fiducial point, this test probes \emph{global} discriminability across the entire hypercube.

\section{Conclusions}
\label{sec:conclusion}
We have introduced a learnable, interpretable generalization of marked statistics for cosmological density fields.
The mark is a translation-equivariant scalar response built from spherical-harmonic-filtered, rotation-invariant channels processed by independent MLPs, so the trained transformation can be opened up and read as a list of morphological preferences before any spectrum is computed.
A contrastive objective aligns the marked summary with the cosmological parameters across the full Latin-hypercube and is residualized against the unmarked~$P_{\delta\delta}$, so the mark is rewarded only for the complementary signal it adds.

On simulated density fields at~$z=0$, the resulting summary improves both Fisher constraints at the fiducial cosmology and held-out parameter recovery over the unmarked~$P_{\delta\delta}$ and classical density marks at~$\kmax\!=\!0.20\,h\,\mathrm{Mpc}^{-1}$, with the largest gains in~$\Omega_m$ and~$\sigma_8$ and a rotation of the joint~$\Omega_m$--$\sigma_8$ contour that breaks the canonical degeneracy.
Mark introspection traces the gain to a small-scale response that recovers signal lost by the unmarked~$P(k)$.
%, while the anisotropic channels remain subdominant in this configuration.
In the latent geometry, the first two principal components of the learned summary embeddings align almost perfectly with~$\sigma_8$ and~$\Omega_m$.

Richer geometrically motivated invariants represent a natural extension to provide more refined interpretability, targeting specific morphological features to better understand the physical origin of the gain and to further improve the summary's constraining power, and we plan to explore this direction in future work.
Furthermore, the same approach could be applied to other set of cosmological parameters such as primordial non-Gaussianity, which inherently requires higher-order statistics to be constrained.

This approach represents a step towards more flexible and interpretable summaries for cosmological inference.

\section*{Acknowledgements}
The authors thank Marco Marinucci, Francesco Spezzati and Nicola Bellomo for useful discussions.
The authors also thank the anonymous reviewers for their constructive feedback, which was useful in refining the presentation of our work.
FS is partly supported by ICSC - Centro Nazionale di Ricerca in High Performance Computing, Big Data and Quantum Computing, funded by European Union - NextGenerationEU.
ML acknowledges support by the MIUR Progetti di Ricerca di Rilevante Interesse Nazionale (PRIN) Bando 2022 - grant 20228RMX4A.
AR acknowledges funding from the Italian Ministry of University and Research (MIUR) through the ``Dipartimenti di eccellenza'' project ``Science of the Universe''.

\section*{Impact Statement}

This work develops scientific machine-learning tools for cosmological inference.
More informative and interpretable summaries can reduce simulation cost and improve the scientific return of galaxy survey data.
We do not identify specific harmful applications.

\bibliographystyle{icml2026}
\bibliography{bibo}

\newpage
\appendix
\onecolumn

\section{Dependence on~$k_{\mathrm{max}}$ and Channel Ablation}
\label{app:kmax_ablation}
\begin{figure}[t]
  \centering
  \includegraphics[width=\textwidth]{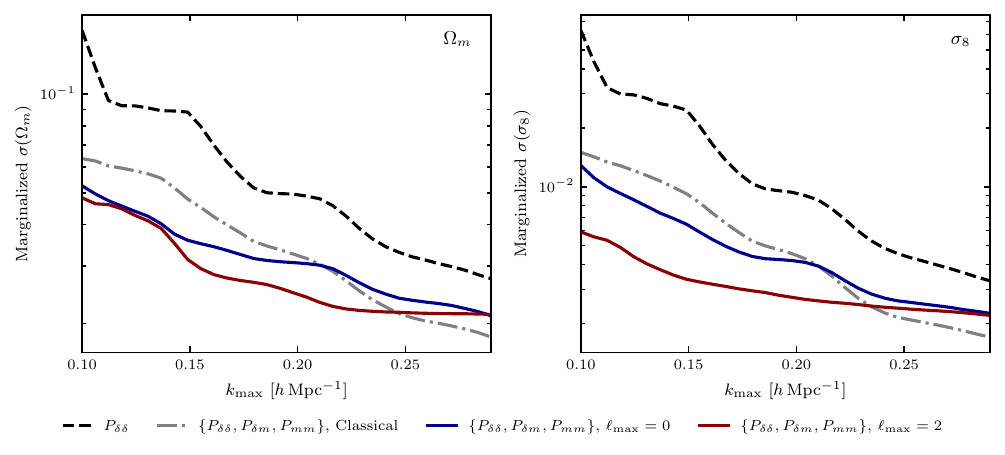}
  \caption{Marginalized~$\Omega_m$ (left) and~$\sigma_8$ (right) constraints as a function of~$\kmax$, for~$\Pdd$ (black), a reference classical mark (gray), the isotropic ablation~$\ell_{\max}=0$ (blue), and the full learned mark~$\ell_{\max}=2$ (red). The Fisher matrix is evaluated for the full set of summaries~$\{\Pdd,\Pdm,\Pmm\}$. The learned marks converge toward the classical baseline as~$\kmax$ grows.}
  \label{fig:kmax_variation}
\end{figure}

In this appendix, we evaluate the robustness of the learned mark to the choice of~$\kmax$ and to explore the contribution of the different~$\ell$-channels to the final summary. 
In \cref{fig:kmax_variation}, we show the marginalized Fisher constraints on~$\Omega_m$ and~$\sigma_8$ as a function of~$\kmax$.
As discussed in the main text, the gain of the learned mark is strongest at~$\kmax\simeq0.2\,h\,\mathrm{Mpc}^{-1}$, where it significantly outperforms both the unmarked power spectrum and standard marking schemes.
At~$\kmax\sim0.3\,h\,\mathrm{Mpc}^{-1}$ the classical marks retain the improvement over~$\Pdd$ while the gain of the learned mark is reduced by the filter taper. 
On the same configuration grid of \cref{fig:results_compact}, at~$\kmax=0.3\,h\,\mathrm{Mpc}^{-1}$, the marginalized ratio~$\sigma_{\mathrm{class}}/\sigma_{\mathrm{learned}}$ spans a range of~$[0.76,1.1]$ for~$\Omega_m$ and~$[0.72,1.18]$ for~$\sigma_8$.

We further explore the contribution of the~$\ell$-channels by comparing the learned mark to an isotropic~$\ell_{\max}=0$ setting.
Keeping the same evaluation pipeline at~$\kmax=0.2\,h\,\mathrm{Mpc}^{-1}$, on the grid of classical marks the isotropic ablation recovers a significant fraction of the MSE ratio in the generalization task, with~$\mathrm{MSE}_{\mathrm{class}}/\mathrm{MSE}_{\mathrm{learned}}\in[0.91,1.33]$ for~$\Omega_m$ and~$\in[0.83,1.34]$ for~$\sigma_8$.
In the Fisher forecast at the fiducial cosmology, the isotropic ablation recovers~$\sigma_{\mathrm{\ell}_{\max}=2}/\sigma_{\mathrm{\ell}_{\max}=0}\sim 0.81$ on~$\Omega_m$ and~$\sim\!0.65$ on~$\sigma_8$ at~$\kmax\simeq0.2\,h\,\mathrm{Mpc}^{-1}$.
A modulation of the local density is therefore responsible for most of the gain, while the anisotropic channels and their cross-interactions provide a refinement that further tightens the final constraining power of the learned mark.
At smaller scales, the effect of the anisotropic channels is reduced by resolution and smoothing effects, and the mark behaves as an effective isotropic modulation of density.

\section{Hyperparameters and Model Structure}
\label{app:hyperparameters}

This appendix collects the hyperparameters used to train the results reported in the main text.
A full training run of the model takes~$\approx\!3$ hours on a single NVIDIA GH200 GPU.

\subsection{Optimization and Contrastive Loss}
\label{app:hp_optim}
The marker network and the summary and parameter embedders are trained jointly with AdamW under the schedule of \cref{tab:hp_optim}.
A short bootstrap phase first pretrains the parameter and~$\Pdd$ projectors under the same contrastive loss (\cref{sec:contrastive}); the~$\Pdd$ projector is then frozen for the main run so that it remains a stable residualization reference.
The objective is an InfoNCE variant that projects the mark summaries, the unmarked~$\Pdd$ summary, and the parameters~$\btheta$ into a shared embedding space, with explicit residualization of the mark embedding against the~$\Pdd$ embedding.
Its projector geometry and negative-sampling scheme are listed in \cref{tab:hp_loss}.

\vspace{1.2em}
\noindent
\begin{minipage}[b]{0.48\textwidth}
\centering
\scalebox{0.8}{
\begin{tabular}{lc}
\toprule
Hyperparameter & Value \\
\midrule
Optimizer                & AdamW \\
Base learning rate       &~$5\times 10^{-4}$ \\
Weight decay             &~$10^{-4}$ \\
Gradient-clip norm       &~$10.0$ \\
Batch size               &~$16$ \\
Mark-parameters LR mult.\      &~$2.0$ \\
$\theta$-proj.\ LR mult.\  &~$0.25$ \\
LR scheduler             & plateau ($\gamma{=}0.5$) \\
patience\,/\,floor &~$8$\,/\,$10^{-7}$ \\
Early-stop patience      &~$20$ ep.\ ($\Delta{=}10^{-5}$) \\
Validation fraction      &~$0.15$ \\
\bottomrule
\end{tabular}
}
\vspace{8pt}
\captionof{table}{Optimization hyperparameters for the contrastive training run.}
\label{tab:hp_optim}
\end{minipage}\hfill
\begin{minipage}[b]{0.48\textwidth}
\centering
\scalebox{0.8}{
\begin{tabular}{lc}
\toprule
Hyperparameter & Value \\
\midrule
Embedding dimension      &~$16$ \\
Mark\,/\,$\Pdd$ proj.\     &~$(\texttt{Lin}\,\texttt{LN}\,\texttt{GELU})^{\times 2}\texttt{Lin}$ \\
Projector hidden width   &~$64$ \\
$\theta$ projector       & linear \\
Temperature~$\tau$       &~$0.05$ \\
$k$-modes   &~$47$~$\left(\kmax\!=\!0.30\,h\,\mathrm{Mpc}^{-1}\right)$ \\
Real negatives\,/\,batch        &~$128$ \\
Synthetic negatives\,/\,batch          &~$384$\,/\,anchor \\
Synthetic close negatives\,/\,batch          &~$32$\,/\,anchor \\
Neighborhood scale       &~$0.15$ \\
Global exclusion scale   &~$0.03$ \\
\bottomrule
\end{tabular}
}
\vspace{8pt}
\captionof{table}{InfoNCE loss and projector heads. Neighborhood and exclusion scales are in normalized~$\theta$.}
\label{tab:hp_loss}
\end{minipage}

\subsection{Network Architectures}
\label{app:marker}
The marker module evaluates the filtered fields of \cref{eq:filtered_fields} in Fourier space, producing~$9$ channels that are reduced to the rotation invariants~$\{E_0,E_1,E_2,I_3\}$, non-linearly processed through independent MLPs and summed together with an additional cross term. 
Its structure is summarized in \cref{tab:hp_mark}.
\label{app:appendix_mlp}
The held-out parameter recovery reported in the main text uses a symmetric-pyramid MLP regressor that maps the marked-summary spectra~$\{\Pdd(k),\Pmm(k),\Pdm(k)\}$ to the cosmological parameters~$\btheta$.
The raw spectra are first processed into three physically motivated channels per~$k$-bin, concatenated into a length-$3N_k$ vector, and fed into the pyramid body, whose architecture settings are listed in \cref{tab:hp_regression}. To train the MLP for parameter regrssion, we use AdamW with Plateau LR scheduler, with early stopping based on the validation loss.

\vspace{1.2em}
\noindent
\begin{minipage}[b]{0.48\textwidth}
\centering
\scalebox{0.8}{
\begin{tabular}{lc}
\toprule
Hyperparameter & Value \\
\midrule
SH channels                  &~$9$ ($\ell{=}0,1,2$) \\
Radial parameterisation      & per-$\ell$ Fourier-feature MLP \\
Radial MLP width      &~$16$ \\
Fourier-feature frequencies  &~$8$ \\
Radial seed~$r_0/r_{\max}$   &~$0.3$ \\
Taper~$k_{\mathrm{pass}}/\kn$  &~$0.85$ \\
Taper floor                  &~$0$ \\
Input transform              &~$\mathrm{sgn}(x)\log(1{+}x)$ \\
Per-invariant MLP            & \texttt{Lin}\,\texttt{GELU}\,\texttt{Lin}\\
MLP width                    &~$16$ \\
Cross-term width             &~$8$ \\
\bottomrule
\end{tabular}
}
\vspace{8pt}
\captionof{table}{Marker module. Per-invariant MLPs use zero-output initialization to set~$M\equiv 1$ as first mark state.}
\label{tab:hp_mark}
\end{minipage}\hfill
\begin{minipage}[b]{0.48\textwidth}
\centering
\scalebox{0.8}{
\begin{tabular}{lc}
\toprule
Hyperparameter & Value \\
\midrule
% \multicolumn{2}{l}{\emph{Architecture}} \\
Input features   &~$\log_{10}\Pdd,\ \log_{10}(\Pmm/\Pdd),\ r_k$ \\
Base width~$B$   &~$3N_k$ \\
Hidden widths    &~$[B,2B,4B,4B,2B,B]$ \\
Block            & \texttt{Lin}\,\texttt{LN}\,\texttt{GELU}\,\texttt{Drop} \\
Dropout          &~$0.1$ \\
Loss             &~$\sum_d \log(\mathrm{MSE}_d+\epsilon)$ \\
\bottomrule
\end{tabular}
}
\vspace{8pt}
\captionof{table}{Symmetric-pyramid MLP regressor for held-out parameter recovery.}
\label{tab:hp_regression}
\end{minipage}

\vspace{1em}

\end{document}